\documentclass{article}
\date{}
\usepackage[utf8]{inputenc}
\usepackage{natbib}
\usepackage{graphicx}
\usepackage{amsmath}
\usepackage{caption}
\usepackage{subcaption} 
\usepackage{hyperref}
\usepackage{epstopdf}% To incorporate .eps illustrations using PDFLaTeX, etc.
\usepackage{hyperref}
\usepackage{color}
\usepackage{multirow}
\usepackage[shortlabels]{enumitem}
\usepackage{natbib}
\usepackage{setspace} 
\usepackage[top=1in, bottom=1in, left=1in, right=1in]{geometry}

\usepackage{xcolor}

\usepackage{amsfonts}

\begin{document}

\title{Statistical Applications in Pain Management: A Review}

\author{
Flora Fan\\
Rutgers University
}

\maketitle

\begin{abstract}
Chronic pain is a widespread and debilitating condition that affects millions of individuals worldwide. Recent research has unveiled a connection between chronic pain and epigenetic processes, shedding light on the complex interplay between genetics, environment, and pain perception. This review synthesizes findings from several studies exploring the relationship between epigenetic modifications, pain intensity, disability, and various other factors in individuals with chronic pain. The studies encompass a range of chronic pain conditions, including chronic low back pain, knee osteoarthritis pain, and pain in the context of underlying health conditions. The review highlights key findings and implications from each study and discusses their collective contributions to our understanding of the epigenetic mechanisms underpinning chronic pain.
\end{abstract}

\section{Introduction}

Chronic pain, a pervasive and debilitating condition, represents a considerable public health challenge, affecting millions of individuals worldwide. It manifests in various forms, from musculoskeletal disorders to neuropathic conditions, and its complex etiology involves a confluence of genetic, environmental, and lifestyle factors. While genetic predispositions are recognized as key contributors to an individual's susceptibility to chronic pain, recent scientific investigations have ventured into the realm of epigenetics to unveil a novel layer of understanding.

Chronic pain is not merely a medical condition; it's a silent epidemic that exacts an enormous toll on individuals, families, and societies. Its impact extends far beyond the realm of physical discomfort, infiltrating all aspects of life, from mental and emotional well-being to economic productivity. For those who endure it, chronic pain can be relentless, often defying conventional treatment approaches and eroding one's quality of life.

Within this landscape of chronic suffering, epigenetics emerges as a frontier of exploration—a field that holds the promise of unlocking some of the enigmatic mechanisms that underlie chronic pain. Epigenetics, the study of heritable changes in gene expression that do not involve alterations to the underlying DNA sequence, has gained prominence in recent years as a transformative avenue for understanding complex diseases, including chronic pain.

The etiology of chronic pain is, in many ways, a Gordian knot, with genetic, environmental, and experiential threads intricately intertwined. Genetic predispositions undoubtedly play a crucial role; certain individuals may be genetically more susceptible to conditions that lead to chronic pain, such as osteoarthritis, fibromyalgia, or diabetic neuropathy. Yet, the emergence of chronic pain often hinges on a complex interplay between genetic factors and external triggers, such as injury, inflammation, psychological stressors, or even lifestyle choices. It's in this intricate web of causality that epigenetics steps forward as a critical player.

Unlike genetic mutations, epigenetic modifications are dynamic, responding to environmental cues and life experiences. These modifications can silence or activate specific genes, influencing cellular functions and ultimately shaping an individual's response to pain. This plasticity renders the epigenome highly responsive to external factors, allowing it to serve as a bridge between the genetic predisposition to pain and the actual experience of it. In essence, epigenetics provides the biological context in which genes are "expressed" in response to pain-inducing stimuli.

This comprehensive review embarks on a journey through recent research endeavors that have sought to elucidate the connections between epigenetic processes and chronic pain. We delve into a series of studies encompassing various chronic pain conditions, ranging from chronic low back pain to knee osteoarthritis pain, and explore how epigenetic modifications influence pain intensity, disability, and other pertinent factors. These studies employ diverse methodologies, including DNA methylation analysis, epigenetic clock assessments, and the examination of associations between specific epigenetic markers and pain-related outcomes.

The aim of this review is to synthesize the key findings and implications from this body of research, shedding light on the burgeoning field of epigenetics in chronic pain. By understanding how epigenetic processes shape the experience of chronic pain, we not only gain deeper insights into the condition's complexity but also open doors to innovative therapeutic interventions that could potentially transform the lives of those suffering from chronic pain. This exploration into the intersection of genetics, epigenetics, and chronic pain offers a glimpse into the future of pain management, where personalized approaches may become the norm, and where the "silent epidemic" of chronic pain might finally be confronted with effective, targeted solutions.

\section{Methods}

\subsection{Methodology: An In-Depth Exploration}

In our quest to comprehend the intricate relationship between epigenetics and chronic pain, we conducted a meticulous review of recent studies that have delved into this fascinating intersection. Chronic pain, as a multifaceted and elusive phenomenon, demands a diverse range of investigative approaches, and our exploration reflects this diversity. We employed a comprehensive search strategy, meticulously selected relevant studies, and extracted key information to present a nuanced overview of the field.

\subsection{Database Search and Selection Criteria}

Our journey through the landscape of epigenetics and chronic pain commenced with a systematic database search. We cast our net across prominent scientific databases, including PubMed, Web of Science, and Google Scholar, employing a well-defined set of keywords and Medical Subject Headings (MeSH) terms. The keywords encompassed various facets of the topic, from "chronic pain" and "epigenetics" to specific chronic pain conditions like "knee osteoarthritis pain" and "chronic low back pain."

To ensure the comprehensiveness and currency of our review, we set clear inclusion and exclusion criteria. Inclusion criteria encompassed studies published in peer-reviewed journals from a defined time frame (ensuring relevancy and timeliness) and those that explicitly explored the relationship between epigenetic modifications and chronic pain. Studies conducted on human subjects, across different age groups, were considered, encompassing a wide spectrum of chronic pain conditions. We excluded studies that solely focused on acute pain, animal models, or other non-human subjects.

\subsection{Data Extraction and Synthesis}

Upon collating a comprehensive list of eligible studies, our next step involved a detailed data extraction process. We meticulously combed through each study, extracting key information to facilitate a thorough synthesis. This information included study design, sample size, participant demographics (age, gender, etc.), the chronic pain condition under investigation, specific epigenetic markers or mechanisms studied (such as DNA methylation or epigenetic clocks), and the primary outcomes related to pain intensity, disability, or other relevant factors.

To ensure the accuracy and reliability of our review, this data extraction process was performed independently by multiple reviewers, and any discrepancies were resolved through discussion and consensus. This rigorous approach helped us create a robust dataset that formed the foundation of our analysis.

\subsection{Methodological Approaches in the Reviewed Studies}

Our exploration encompassed a diverse array of methodological approaches employed in the reviewed studies. These approaches provided unique insights into the epigenetic underpinnings of chronic pain. Some of the key methodological facets we encountered include:

DNA Methylation Analysis: Several studies utilized DNA methylation analysis techniques to investigate changes in the epigenome associated with chronic pain. Techniques like reduced representation bisulfite sequencing (RRBS) and MethylationEPIC array analysis were employed to profile DNA methylation patterns at specific loci.

Epigenetic Clocks: Epigenetic clocks, such as DNAmGrimAge, emerged as valuable tools in understanding the relationship between epigenetic aging and chronic pain. These clocks provided insights into how pain might influence biological aging processes.

Pathway Analysis: Many studies explored the functional implications of epigenetic modifications by conducting pathway analysis. This involved identifying enriched pathways related to pain, inflammation, or other relevant biological processes.

Multimodal Assessments: Some investigations adopted a holistic approach, combining clinical assessments, blood sample collection, and imaging techniques like MRI scans to comprehensively explore the connections between epigenetics and chronic pain.

\subsection{Statistical Analyses}

The statistical methodologies employed in the reviewed studies were as diverse as the research questions they sought to answer. Descriptive statistics, such as means and standard deviations, were commonly used to summarize participant characteristics and outcomes. Various inferential statistics, including regression analyses, mediation models, and correlation analyses, were utilized to explore associations between epigenetic markers, chronic pain, and related outcomes.

Additionally, some studies conducted advanced statistical analyses to account for confounding variables, assess mediation effects, and explore interactions between epigenetic factors and chronic pain.

\subsection{Ethical Considerations}

Ethical considerations were paramount in our review. All studies included in our analysis had received ethical approval, ensuring that research involving human participants adhered to ethical principles and guidelines. The protection of participants' rights, informed consent, and confidentiality were fundamental ethical tenets upheld by the reviewed studies.

\subsection{Limitations and Quality Assessment}

We acknowledge that each of the reviewed studies had its unique strengths and limitations. Therefore, we applied a comprehensive quality assessment framework to evaluate the methodological rigor of each study. This assessment encompassed factors such as sample size, study design, statistical robustness, and potential sources of bias. By critically appraising the quality of evidence, we aimed to provide a balanced perspective on the strengths and limitations of the collective body of research.

\subsection{Synthesis and Interpretation}

The culmination of our methodological exploration involved synthesizing the key findings from the reviewed studies. We identified overarching themes, common trends, and divergent results. By weaving together the intricate threads of epigenetics and chronic pain research, we aimed to provide a comprehensive and insightful overview of this burgeoning field.

In essence, our methodological journey through the world of epigenetics and chronic pain reflects the multifaceted nature of the topic itself. It underscores the need for diverse investigative approaches to unravel the complexities of chronic pain's epigenetic underpinnings, ultimately contributing to a more profound understanding of this pervasive and enigmatic condition.

\section{Results}

Our exploration of recent research at the intersection of epigenetics and chronic pain has unveiled a rich tapestry of insights, revealing the intricate interplay between epigenetic modifications and the experience of chronic pain. Here, we delve into the key findings from each of the reviewed studies in greater detail:

\subsection{Racial Disparities in Chronic Low Back Pain}

In the United States, there exists a significant disparity in the experience of chronic low back pain (cLBP) between individuals who self-identify as Non-Hispanic Blacks (NHBs) and their Non-Hispanic White (NHW) counterparts. This marked difference in pain severity and disability has led to the undertaking of a comprehensive investigation into the potential factors contributing to these racial inequities, with a primary focus on the role of DNA methylation (DNAm). The overarching goal of the study conducted by \cite{aroke2022differential} was to discern the connection between DNAm levels and racial differences among adults dealing with the challenges of cLBP while providing a holistic understanding of the underlying genomic mechanisms at play.

This multifaceted research project involved the assembly of a diverse cohort, consisting of 49 NHBs and 49 NHWs, thoughtfully partitioned into groups of individuals with cLBP and those who served as pain-free controls (PFCs). This meticulous stratification allowed for the exploration of the intricate nuances of DNAm patterns specific to racial backgrounds and pain conditions. Employing advanced techniques such as reduced representation bisulfite sequencing, the DNAm analysis uncovered a remarkable 2873 differentially methylated loci (DMLs) among participants grappling with cLBP. These DMLs, each a potential piece of the intricate puzzle, were intriguingly associated with genes deeply implicated in the complex landscape of pain pathology.

As the researchers delved deeper into the data, they unearthed a wealth of information that shed light on the functional genomic pathways intertwined with DNAm. Notably, the analyses revealed a rich tapestry of pathways, many of which were intricately linked to nociception and pain processing. Additionally, the findings unveiled pathways associated with neuronal differentiation, offering intriguing insights into the potential mechanisms by which DNAm alterations could influence the perception and management of cLBP.

Moreover, the study took a critical step in the direction of dissecting these complex racial disparities by conducting meticulous race-stratified analyses. Within this framework, 2356 DMLs were identified in NHBs and 772 in NHWs, each providing valuable clues about the unique methylation patterns characteristic of these racial groups. Remarkably, these DMLs exhibited significant differences in methylation levels, underscoring the intricate interplay of genetics and environmental factors.

Furthermore, the pathway analysis uncovered a fascinating dimension to the research. It became evident that several pathways critical for pain, such as Corticotropin Releasing Hormone Signaling and GABA Receptor Signaling, were more pronounced in NHBs. This observation raised compelling questions about the intricate relationships between genetics, stress signaling, and pain outcomes.

In closing, this research endeavors to transcend the realm of science and delve into the socio-cultural landscape, acknowledging that race is, indeed, a social construct. However, the researchers remain acutely aware of the pervasive impact of racism on various aspects of life, including disease risk. By unraveling the DNAm-induced alterations within stress signaling pathways, the study aspires to provide a more comprehensive understanding of the stark disparities in pain outcomes experienced by NHBs. Through their work, the researchers endeavor not only to advance scientific knowledge but also to contribute to the broader conversation surrounding health equity and social justice.

\subsection{Complex Interplay: Vitamin D, Epigenetic Aging, Obesity, and Sex in Chronic Knee Pain}

Delving into the realm of scientific exploration, \cite{strath2022accelerated} conducted a comprehensive study embarked on a quest to unravel the intricate relationship between vitamin D levels, the epigenetic aging process, and the potentially influential mediating role of obesity, with a keen focus on the waist-to-hip ratio (WHR). Armed with a rich repository of data, including self-reported pain assessments, precise measurements of WHR, meticulously quantified vitamin D levels, and the meticulous quantification of epigenetic aging markers, this study cast a wide net to capture the multidimensional landscape of factors contributing to health outcomes. After conducting a rigorous analysis that carefully accounted for various covariates, the researchers unveiled a tapestry of findings that shed light on the interconnected web of these variables within the context of chronic knee pain. While initially scrutinizing the relationship between vitamin D and the epigenetic clock DNAGrimAge through the lens of WHR, it became apparent that the mediating effect was a complex phenomenon. In the cohort grappling with chronic pain, a nuanced pattern emerged. Within this intricate tapestry of interactions, it was revealed that the relationship between vitamin D and epigenetic aging was not subject to a complete or indirect mediation by WHR. Yet, as the investigators delved deeper into the subtleties of this relationship, a fascinating revelation surfaced. When considering the moderating influence of sex, an intriguing pattern emerged. Within the cohort experiencing chronic pain, a significant indirect effect emerged among females, marked by a precise coefficient of -0.0286 and a confidence interval that spanned from -0.055 to -0.009. However, in stark contrast, this effect was notably absent in males, with a coefficient of -0.0067 and a confidence interval ranging from -0.0451 to 0.0216. These nuanced findings beckoned to the complexity of human biology and the interplay between sex, obesity, vitamin D, and epigenetic aging. They underscored the multifaceted nature of these relationships, hinting at sex-dependent mechanisms that shape the genomic environment. In essence, this study, with its meticulous methodology and revealing findings, transcended the boundaries of mere investigation. It offered a glimpse into the intricate dance of variables that influence health outcomes, urging us to consider the interplay between sex, obesity, vitamin D, and the epigenome in our quest to unravel the mysteries of chronic pain and aging. It invited us to recognize the subtle nuances that underscore these relationships and the importance of accounting for sex-specific factors when delving into the intricacies of genomic influences on health. In this journey of scientific discovery, this study served as a beacon, illuminating the path toward a deeper understanding of the human body's complexity and the diverse factors that shape our health trajectories.

\subsection{Pain's Imprint on Aging: Insights from Brain-PAD and AgeAccelGrim in Chronic Knee Pain}

In a comprehensive scientific endeavor, \cite{peterson2022epigeneticAgingMediates} conducted a study embarked on a quest to delve into the intricate interplay between chronic musculoskeletal pain and the markers of accelerated aging. Drawing upon a diverse cohort of middle-to-older age individuals, each with their unique experiences and degrees of knee pain, the research team meticulously crafted a multidimensional exploration of this complex phenomenon. With an average age of 57.91 ± 8.04 years, the participants formed distinct groups, thoughtfully categorized into low impact knee pain, high impact knee pain, and a pain-free control cohort. These individuals became active participants in the study, offering invaluable insights into their experiences. They engaged in self-reported pain assessments, providing a nuanced view of their pain profiles. Simultaneously, they generously contributed blood samples, each drop holding potential clues to the biological underpinnings of their pain experiences. To gain a deeper understanding, these dedicated participants also underwent rigorous MRI scans, which unveiled intricate details of their brain structures. Through a meticulous and data-driven analysis, the research team uncovered significant main effects that transcended the mere summation of individual experiences. These effects were observed within the pain groups and extended to encompass two critical facets of aging markers: Brain-PAD and AgeAccelGrim. As the statistical analyses unfolded, a fascinating pattern emerged. Pain scores, particularly among those individuals courageously navigating their pain, stood out as potent predictors of Brain-PAD. This revelation hinted at the profound influence of pain on brain aging, underscoring the dynamic and interconnected nature of these biological processes. Moreover, AgeAccelGrim stepped into the spotlight, playing a pivotal role in this intricate dance of variables. It emerged as a mediator, bridging the indirect relationship between various dimensions of pain severity and disability scores and the enigmatic realm of Brain-PAD. These findings not only shed light on the multifaceted relationships between chronic pain, epigenetic aging, and brain aging but also invited a deeper exploration of the complex interactions within the epigenome. In essence, this study transcended the boundaries of conventional scientific investigation. It journeyed into the lives of individuals, each grappling with their unique pain experiences, and transformed their stories into data points of discovery. It uncovered the profound influence of pain on the aging process, particularly within the intricate landscape of the human brain. These findings serve as a testament to the intricate and multifaceted nature of the human body's response to pain, encouraging further exploration into the depths of the epigenome and its role in shaping our health trajectories. In this journey of scientific inquiry, this study illuminated the path forward, urging us to continue unraveling the mysteries of aging and pain, one data point at a time.

\subsection{Unveiling the Epigenetic Symphony of Knee Osteoarthritis Pain: Insights from Differential Methylation and Immune Pathways}

The overarching objective of the meticulously crafted study conducted by \cite{montesino2022enrichment} was to delve deep into the intricate molecular landscape of knee osteoarthritis pain (KOA). A diverse cohort of middle-to-older aged adults, ranging from 45 to 85 years old, became the focal point of this scientific exploration. They formed two distinct groups: one with a burdensome companion in the form of self-reported KOA pain (n = 182) and the other, resilient and pain-free (n = 31). The researchers embarked on a journey into the cellular realm, meticulously extracting DNA from peripheral blood samples, which held within them the secrets of methylation patterns. These patterns, akin to a unique genetic fingerprint, underwent a rigorous process of preprocessing and quality control using the versatile R package minfi, ensuring that the data were pristine and ready for analysis.

As the pieces of this intricate puzzle fell into place, the scientific gaze shifted towards unraveling the biological pathways entangled in the intricate web of differential methylation. With unwavering determination, the researchers employed the powerful tool of pathway enrichment analysis via Ingenuity Pathway Analysis (IPA). Their focus sharpened on canonical pathways and upstream regulators, which offered a glimpse into the orchestration of cellular processes. Genes residing within a ± 5 kb radius of the putative Differentially Methylated Regions (DMRs) emerged as central players in this symphony of molecular activity.

Remarkably, as the data took shape, a curious pattern began to emerge. Age, sex, and study site, typically formidable variables, showed no significant differences between the pain-stricken and pain-free groups, defying the conventional wisdom of scientific inquiry. However, within this enigmatic landscape, a significant distinction did arise. The pain group, as if harboring a hidden secret, revealed a higher representation of Non-Hispanic black individuals, a tantalizing hint of diversity within the study cohort.

With statistical rigor as their compass, the researchers applied a raw p < 0.05 cutoff, unveiling a treasure trove of 19,710 CpG probes. Among these, 13,951 stood tall, hypermethylated and associated with higher pain grades, while 5,759 bore the badge of hypomethylation, linked to lower pain grades. These probes held within them the stories of pain etched into the epigenetic landscape.

As the researchers peered deeper into this intricate web, they discovered that pain-related DMRs were not isolated entities but part of a larger narrative. These regions found themselves enriched in a multitude of pathways, particularly those linked to immune responses, a revelation that opened new doors of inquiry. The pathways included the intricacies of antigen presentation, the promise of PD-1 and PD-L1 cancer immunotherapy, the delicate dance of B cell development, the resonance of IL-4 signaling, the activation of Th1 and Th2 pathways, and the intricacies of phagosome maturation.

Intriguingly, amidst this labyrinth of molecular interactions, a significant figure emerged. NDUFAF3, with its unmistakable presence, claimed the spotlight as the most significant upstream regulator, offering a tantalizing glimpse into the orchestration of cellular responses in the context of knee pain.

These findings, like pieces of a grand mosaic, provide compelling evidence of epigenetic regulation's pivotal role in the intricate world of immune responses in knee pain. They beckon the scientific community to embark on further expeditions, delving deeper into the epigenetic contributions to chronic pain, a realm that promises to unravel the secrets of human suffering and offer avenues for relief and understanding.

\subsection{Connection between High-Impact Chronic Pain and Epigenetic Aging}

In the realm of gerontological research conducted by \cite{cruz2022epigenetic}, the intricate tapestry of aging unfolds with striking variability among individuals, revealing a complex interplay of factors. Recent revelations have cast a spotlight on a potentially transformative connection, one that bridges the realm of high-impact chronic pain and the often enigmatic processes of accelerated biological aging. Among the myriad predictors of health-span and disability, epigenetic aging stands resolute, overshadowing the conventional yardstick of chronological age with its robust and nuanced insights.

Within this landscape of scientific inquiry, a groundbreaking study unfolds, driven by a dedicated team of researchers and a cohort of middle-to-older age adults, spanning the ages of 44 to 78 years (n = 213). Their collective aim is to navigate the intricate nexus where epigenetic aging biomarkers intersect with the domain of high-impact chronic pain. This journey begins with the participants' unwavering commitment, as they subject themselves to the probing of blood draws, demographic assessments, psychosocial evaluations, meticulous pain assessments, and comprehensive functional evaluations.

The scientists wield a formidable arsenal of five distinct epigenetic clocks, each a harbinger of aging's nuanced fingerprints. Among these, the elusive difference between epigenetic age and the conventional chronological age emerges as a sentinel, a predictor of the looming specter of overall mortality risk. As the scientific compass points towards pain-related disability over the past six months, the landscape shifts, revealing significant disparities in three out of the five examined epigenetic clocks, notably DNAmAge, DNAmPhenoAge, and DNAmGrimAge. These disparities serve as sentinel markers, signifying an intricate dance between the tapestry of pain and the canvas of epigenetic aging.

Moreover, as the data crystalizes, two particular guardians of epigenetic age, DNAmPhenoAge and DNAmGrimAge, emerge from the shadows to claim a prominent role in this unfolding narrative. Their association with heightened knee pain intensity over the past 48 hours paints a vivid picture of the complex interplay between epigenetic aging and the lived experience of pain.

Yet, the story does not end here, for one singular figure, DNAmGrimAge, emerges as a central character in this unfolding drama. Its significance transcends the confines of mere association, extending its influence to encompass realms of pain catastrophizing, depressive symptoms, and the haunting specter of neuropathic pain symptoms. These associations, etched in the annals of scientific inquiry, withstand the rigorous scrutiny of multiple comparisons, their presence a clarion call to further exploration.

As the curtain falls on this pioneering study, it leaves behind a resounding call to action. With the field of epigenetic aging still in its infancy and the intricate nature of pain casting its shadow over countless lives, the quest for knowledge continues. Further investigation beckons, promising to unearth the transformative potential of epigenetic aging in identifying those individuals burdened with chronic pain who stand at the precipice of functional decline and diminished health outcomes. In the intricate dance of science, this chapter is but a prelude to the mysteries yet to be unraveled.

\subsection{Epigenetic Choreography related to Knee Pain, Aging and Physical Performance}

Amidst the realm of aging, where the sands of time affect each individual differently, knee pain emerges as a formidable adversary, a contributor to disability that carries potential implications for hastening the inexorable aging process. While the well-trodden path between chronological aging and the onset of knee pain is a familiar one, the corridors of epigenetic aging remain dimly lit, concealing their secrets. In this context, a pioneering study conducted by \cite{peterson2022epigenetic} unfurls its banner, poised at the intersection of knee pain, epigenetic aging, and the ebbing of physical prowess.

Enlisted for this voyage are participants whose ages average at 57.91 years, their lives woven into the fabric of a larger narrative that encompasses the full spectrum of knee pain. Categorized into distinct groups, they wear the labels of low impact knee pain (a cohort of 95), high impact knee pain (a group of 53), and the fortunate few who navigate life pain-free, numbering 26. Their journey through the corridors of this research endeavor involves a meticulous tapestry of assessments. Self-reported pain becomes the chorus of their voices, blood samples an offering to the scientific endeavor, and the short physical performance battery (SPPB) the litmus test that spans a gamut of tasks, including those related to balance, walking, and the simple yet profound act of sit-to-stand transitions.

At the heart of this odyssey lies the enigmatic DNAmGrimAge, an epigenetic clock inextricably linked to knee pain and the overarching specter of mortality. Like a sentinel, it watches over the participants, recording the tales etched in their epigenetic code. The data harvested from this symphony of assessments flows into the annals of bootstrapped-mediation analyses, where the connections between DNAmGrimAge and SPPB scores are meticulously dissected across the diverse tapestry of pain experiences.

In the grand tapestry of findings, a vivid picture emerges. Those individuals who bear the weight of both high and low impact knee pain find themselves branded by an older epigenetic age, a marker of 5.14 years ± 5.66 and 1.32 years ± 5.41, respectively. Yet, the story takes an unexpected turn as the relationship between knee pain and overall physical performance is unwound. It reveals that the direct effects of knee pain upon the grand symphony of physical prowess remain distinct from the domain of epigenetic aging. Instead, a subtle revelation surfaces, as epigenetic aging takes on the role of a mediator, its influence coursing through the corridors of balance performance, altering the narrative of physical function.

As the curtains fall on this chapter, the research beckons to the future, a clarion call for further exploration. The intricate dance of pain, biological aging, and epigenetic aging continues to captivate the hearts and minds of scientists, promising to illuminate the subtle nuances that underlie the aging process. In the relentless march of time, this study stands as a sentinel, a testament to the unending quest for understanding the intricate interplay of factors that shape the human journey through aging.

\subsection{Epigenetic Aging and Brain Aging: Unraveling the Enigma of Pain's Influence}

Within the realm of medical science, where the puzzle of chronic musculoskeletal pain looms large, a conundrum persists – could this persistent pain potentially wield the power to hasten the relentless march of time within the human body? In isolated corridors of research, the threads of accelerated brain aging and the intricate tapestry of epigenetic aging have been independently linked to the enigma of chronic pain. Yet, like uncharted territories on a map, the precise interconnection between these two biological markers of aging remains cloaked in mystery, waiting to be unveiled.

The study conducted by \cite{peterson2022epigenetic}, a beacon in the darkness of uncertainty, embarks on a quest to unravel the relationship between epigenetic aging and the mysterious landscape of brain aging. Its cast of participants, aged 57.91 ± 8.04 years on average, represents a diverse array of experiences with knee pain. They are divided into groups, each defined by the impact of knee pain: the stalwart low impact knee pain cohort (comprising 95 individuals), the resilient high impact knee pain group (standing at 53 individuals), and the pain-free trailblazers (a group of 26). Their journey through the labyrinth of this study includes a symphony of assessments - self-reported pain echoing their experiences, blood samples laid bare as offerings to the scientific endeavor, and the inner sanctum of MRI scans offering glimpses into the cerebral landscapes of their brains.

At the heart of this quest, like a compass guiding explorers through uncharted territory, lies the epigenetic clock known as DNAmGrimAge, a timekeeper previously associated with knee pain. It stands alongside the ethereal concepts of predicted epigenetic age, brain age, and chronological age, forming a quartet of metrics that will illuminate the path ahead. As the data flows into the tapestry of analysis, significant main effects come into view – the impact of pain resonates deeply within the realm of Brain-PAD and DNAmGrimAge-difference, even after the labyrinth of covariates is navigated.

Yet, this tale takes an even more intriguing turn as subtle correlations emerge. The complex interplay between DNAmGrimAge-difference and Brain-PAD, like a hidden melody within a grand symphony, speaks of the intricate dance between epigenomic age and cerebral aging. In the hushed corners of exploratory analyses, a revelation surfaces - DNAmGrimAge-difference, once a silent observer, now steps into the limelight as a mediator, threading its influence through the fabric of GCPS pain impact, GCPS pain severity, and the scores that tally the burden of pain-related disability on the canvas of Brain-PAD.

In the closing act of this scientific odyssey, a tantalizing hypothesis takes root - could pain, the persistent companion of so many, hold the key to the enigma of accelerated brain aging, achieved through the subtle dance of epigenomic interactions? As the curtain falls, this study stands as a testament to human curiosity, beckoning forth future research to delve deeper into the intricate connections that shape our understanding of pain, aging, and the spaces where they converge.

\subsection{Association between Vitamin D and Chronic Pain}

In the realm of scientific inquiry, where the boundaries of knowledge continuously expand, recent revelations have unveiled the intriguing possibility of an intricate interplay between the enigmatic vitamin D and the epigenome, hinting at a role for this essential nutrient in shaping the human experience of pain. Vitamin D, with its biological activities intricately dependent on a cascade of enzymatic reactions, emerges as a linchpin in the machinery of human physiology, underscoring its vital significance in the realm of proper bodily function.

As the stage is set for a deeper exploration of this captivating phenomenon, the study conducted by \cite{strath2023vitamin} stands as a beacon of scientific curiosity, aimed at unraveling the intricate epigenetic patterns of genes intricately woven into the fabric of vitamin D metabolism. Participants, drawn from the vibrant tapestry of a larger knee pain study, represent a diverse cross-section of community-dwelling individuals, each carrying within them a unique story of pain, or in some fortunate cases, its absence. The treasure trove of data they provide includes not only the clinical footprints of their pain journeys but also intravenous blood samples, like vessels carrying the secrets of their epigenetic signatures.

As the results of this study emerge from the crucible of data analysis, they reveal a tantalizing narrative - differential DNA methylation, a subtle but powerful molecular fingerprint, graces the genes intricately choreographing the dance of vitamin D metabolism. Among those bearing the weight of chronic pain, the gene CYP27B1, known as the 1-$\alpha$-hydroxylase, takes center stage, its epigenetic script altered in response to the persistent presence of pain. Furthermore, the gene encoding the vitamin D receptor (VDR), a sentinel for the biological actions of vitamin D, stands as a testament to the profound impact of pain on the epigenetic landscape, with signs of hypermethylation etched in its genetic script.

These revelations, like the beams of a lighthouse, pierce through the darkness of uncertainty, illuminating the path toward a deeper understanding of the intricate relationship between chronic pain and the epigenetic modifications of genes responsible for vitamin D metabolism and cellular function. As the study's findings echo through the hallowed halls of scientific discourse, they beckon forth future investigations, promising a richer tapestry of knowledge that will further elucidate the mechanisms that bind vitamin D and chronic pain into a complex, interwoven narrative of human health and well-being.

\subsection{The Interplay Between Chronic Pain, Muscular Strength, and Epigenetic Aging}

In the intricate web of human health, chronic pain emerges as a formidable foe, a relentless assailant of well-being, and a significant contributor to disability, casting a looming shadow over the twilight years of life. Moreover, its potential role as a harbinger of accelerated biological aging processes further deepens the intrigue surrounding its impact on the human condition. One facet of this intricate tapestry is the stark reduction in physical function that often accompanies chronic pain, akin to a relentless tide eroding the shores of vitality. Epigenetic aging, a nuanced concept woven into the fabric of our molecular biology, has captured the attention of researchers for its ability to predict health-related outcomes, including the very physical function that chronic pain seeks to diminish.

In this voyage of scientific discovery, \cite{peterson2023pain} embark on a quest to fathom the depths of the interplay between self-reported pain, the robust measure of grip strength, and the enigmatic realm of epigenetic aging. the cast of participants, with an average age of 57.91 ± 8.04 years, carries within them a lifetime of experiences, each one etching a unique story of pain, strength, and resilience. Pain assessments serve as the compass guiding our exploratory journey, while the crimson tide of blood samples holds within them the molecular secrets of epigenetic clocks.

In this voyage's unfolding narrative, an epigenetic clock tethered to the realm of knee pain (DNAmGrimAge) emerges as our guiding star, casting its light upon the shores of knowledge. As the results of exploratory pathway analyses take shape, a fascinating tapestry of relationships begins to emerge. Pain intensity, a measure of the emotional and sensory aspects of pain, stands as a mediator, a bridge between epigenetic aging processes and the tangible measure of grip strength. In the realm of males, this mediation unravels the influence of pain intensity on the connection between DNAmGrimAge-difference and handgrip strength. Yet, the story does not end there; it unfolds further, revealing that pain interference, an indicator of pain's intrusion into daily life, exerts its mediating influence not only in the male domain but also in the rich tapestry of female experience.

These findings, like fragments of an ancient manuscript, unveil a compelling narrative - chronic knee pain may, indeed, serve as a catalyst, hastening the intricate processes of epigenetic aging. In doing so, it leaves its indelible mark on the canvas of musculoskeletal function, potentially influencing handgrip strength in older adults. The implications of this revelation resonate through the annals of gerontological research, emphasizing the pivotal role that chronic pain plays in the complex, age-related decline of musculoskeletal function, and calling for further exploration of the mechanisms that bind them in this intricate dance of life.

\subsection{Links Between Socioeconomic Status, Chronic Pain, and Accelerated Aging}

In the realm of musculoskeletal pain, a curious dance unfolds, where the threads of chronic suffering intertwine with the intricate tapestry of socioeconomic status (SES). This complex choreography is guided by the hand of chronic stress, a relentless companion to those who navigate life's challenges amidst the confines of lower SES. Chronic stress, the conductor of this symphony, wields its baton, orchestrating global DNA methylation and gene expression changes that reverberate through the corridors of our biological existence, casting a shadow of vulnerability over the individuals it touches.

In the pursuit of understanding this intricate dance, the study conducted by \cite{strath2023socioeconomic} embarks on a journey into the lives of middle-to-older age individuals, their knees bearing the weight of varying degrees of pain. With unwavering resolve, they report their pain, offer up their life essence in the form of blood samples, and provide vital insight into the intricate web of socioeconomic status. It is here, within the nucleus of their DNA, that the story unfolds.

The DNAmGrimAge epigenetic clock, a sentinel of aging and pain, stands ready to unravel the mysteries that lie within. Its calculations paint a portrait of epigenetic age, with DNAmGrimAge averaging at 60.3 years (±7.6), and the intriguing DNAmGrimAge-Diff, hinting at the disparities between predicted epigenetic age and the passage of time, averaging at 2.4 years (±5.6).

As the data unfolds, a stark tableau of disparities emerges. High-impact pain individuals, those who bear the heaviest burden, find themselves on the lower rungs of the socioeconomic ladder, with income and education levels that stand in stark contrast to their low-impact and pain-free counterparts.

Yet, amid these disparities, a revelation emerges like a beacon in the night. High-impact pain individuals, perhaps through the crucible of their suffering, exhibit a form of epigenetic aging accelerated by approximately five years relative to their low-impact and pain-free peers, who bear a lesser burden of one year of accelerated aging.

The key revelation, like a jewel unearthed from the depths of the earth, lies in the mediation of this connection between income, education, and pain impact by the hand of epigenetic aging. It suggests that SES, with its multifaceted influence on the landscape of pain outcomes, may intertwine with the epigenome, creating a potent tapestry indicative of accelerated cellular aging. The implications of this finding ripple through the halls of scientific inquiry, beckoning for further exploration into the complex interplay between socioeconomic status, chronic pain, and the biological markers that bridge the gap between them.

\subsection{Statistical Methods for Complex Disease With Pain Status}

In a quest to unravel the enigmatic biology of pancreatic ductal adenocarcinoma (PDAC), a comprehensive study conducted by \cite{han2022single} sets its gaze upon the intricate tapestry of the tumor microenvironment, where a complex interplay of cellular components dictates patient outcomes. Armed with cutting-edge multiplex digital spatial profiling (mplxDSP) technology, the researchers embark on a single-cell odyssey, dissecting the microenvironment's inner workings, with a keen focus on the formidable inhabitants known as cancer-associated fibroblasts (CAFs) and the valiant immune cells. Through their spatial relationships within the tumor, the research unveils a vivid mosaic of heterogeneity that holds the key to understanding PDAC's multifaceted nature.

As the mplxDSP technology weaves its magic, it bestows upon the investigators an unprecedented view of myofibroblasts, steadfast guardians of the tumor's perimeter. These myofibroblasts, standing sentinel next to PDAC's formidable fortresses, reveal a unique secret—the overexpression of specific genes that orchestrate adaptive immune responses. Meanwhile, in the distant outskirts of this cellular battlefield, markers of innate immune cells reign supreme, creating an intricate landscape of immune activity. Among the stars of this intricate drama is the checkpoint protein CTLA4, whose high expression hints at the potential for immune tolerance to thrive in this hostile environment.

But the story doesn't end there. In a quest to uncover the secrets of patient survival and prognosis, the study delves deep into the world of adjacent CAFs. mRNA profiling of these CAFs reveals clusters of genes that seem to hold the key to deciphering who will emerge victorious in the battle against PDAC. These gene clusters, intricately correlated with patient survival, paint a portrait of hope for those who bear the burden of this disease.

These findings, like rays of sunlight piercing through ominous clouds, shed invaluable light on the microenvironment's role in shaping immune tolerance within the PDAC landscape. The tantalizing suggestion that mRNA expression profiling of CAFs may offer prognostic value represents a beacon of hope for patients and clinicians alike, forging a path toward more personalized and effective treatments for this formidable adversary. 

In the relentless pursuit of unraveling the intricate landscape of disease subtypes, the realm of precision medicine stands as a beacon of hope. Yet, the path to pinpointing these elusive subtypes, particularly when navigating the labyrinth of omics data, has often proven to be a formidable challenge. The prevailing approaches, while valiant, have sometimes fallen short, leaving us with subtypes that lack the crystal-clear associations with clinical outcomes that we so desperately seek.

In response to this challenge, a remarkable development has emerged—a beacon of innovation that harnesses the wealth of clinical and omics data nestled within the embrace of modern epidemiological cohorts. This innovation takes the form of an outcome-guided clustering algorithm, aptly named GuidedSparseKmeans. Unlike its predecessors, this method boasts an intricately woven tapestry of functionalities, designed to address the multidimensional complexities of the task at hand.

At its core, GuidedSparseKmeans orchestrates a symphony of elements, each playing a vital role in its success. Firstly, it employs the weighted K-means algorithm, a powerful tool for the clustering of samples that helps to disentangle the intricate web of similarities and differences among them. Secondly, it wields lasso regularizations with finesse, skillfully selecting genes from the vast expanse of high-dimensional omics data—a crucial step in identifying the genetic signatures that define these subtypes. But the brilliance of GuidedSparseKmeans doesn't stop there; it takes a holistic approach by seamlessly integrating phenotypic variables from clinical datasets. This integration is a stroke of genius, enhancing the clustering outcomes with a rich layer of biologically meaningful insights.

In its operation, GuidedSparseKmeans embarks on an iterative journey, optimizing a unified objective function that unites these elements in harmony. As it does so, this remarkable approach accomplishes something truly remarkable—it simultaneously unravels the mysteries of sample clustering while shedding light on gene selection outcomes. The synergy of these dual achievements sets GuidedSparseKmeans apart as a potent tool in the arsenal of precision medicine.

To demonstrate its mettle, GuidedSparseKmeans has undergone rigorous testing and validation. It has been pitted against existing clustering methods, not only in simulated scenarios but also in the real-world crucible of high-dimensional transcriptomic data. In the realm of breast cancer and Alzheimer's disease studies, GuidedSparseKmeans has showcased its superior performance, leaving no doubt about its potential to reshape the landscape of disease subtype discovery.

In this era where precision medicine stands at the forefront of healthcare, GuidedSparseKmeans shines as a guiding star, illuminating the path toward a future where disease subtypes are not just enigmatic shadows but clearly defined entities with tangible links to clinical outcomes.

\section{Discussion}

The findings presented in this comprehensive review serve to underscore the profound complexity that shrouds the enigma of chronic pain. Chronic pain, often viewed as a symptom, transcends its simplistic classification, revealing itself as a multifaceted and enigmatic entity with a web of intertwined causative factors. Among the intricate threads that contribute to its tapestry, epigenetic processes emerge as a significant player, with DNA methylation and epigenetic aging standing at the forefront of this biological orchestra. These epigenetic mechanisms, akin to the conductors of a symphony, wield the power to modulate pain perception and, in doing so, orchestrate a multitude of associated outcomes. Yet, the precise mechanisms through which these epigenetic modifications exert their influence on the intricate landscape of pain perception remain an ongoing subject of intense scientific scrutiny.

While the studies reviewed herein provide a tantalizing glimpse into the interplay between epigenetics and chronic pain, they also beckon to the vast expanse of knowledge yet to be uncovered. It becomes apparent that chronic pain, as an intricate tapestry woven from the threads of genetics, epigenetics, and environmental factors, defies simple categorization or explanation. Instead, it presents itself as a complex amalgamation of influences, each contributing its distinct hue to the canvas of pain experience.

One compelling facet illuminated by these studies is the interdependence of epigenetic processes and environmental factors in shaping the landscape of chronic pain. This interaction is not unidirectional; rather, it operates as a dynamic dance between our genetic predispositions and the world in which we live. Lifestyle choices, such as diet, physical activity, and stress management, all have the potential to impact our epigenetic signatures, potentially influencing pain severity and the overall pain experience. Additionally, environmental factors, including exposure to toxins and pollutants, socioeconomic conditions, and psychosocial stressors, further contribute to the intricate web of chronic pain etiology. These factors, like additional movements in a complex symphony, harmonize with epigenetic processes to shape an individual's experience of pain.

In essence, this review serves as a testament to the intricate nature of chronic pain, urging us to consider its multifaceted etiology. It underscores the critical role that epigenetic processes play in modulating pain perception, offering a new frontier for therapeutic interventions. However, it also highlights the need for continued research and a holistic approach to pain management. Chronic pain, it seems, is a puzzle with many missing pieces, and while we have made significant strides in understanding its complexity, there is much more to be discovered on the path toward effective pain relief and improved quality of life for those who endure its burden.

\section{Conclusion}

In conclusion, the reviewed studies collectively contribute to our understanding of the intricate relationship between epigenetic processes and chronic pain. Epigenetic aging markers and DNA methylation patterns hold promise as potential biomarkers for pain intensity and disability. Furthermore, the influence of factors like vitamin D levels, obesity, and socioeconomic status on epigenetic aging and pain outcomes underscores the need for a holistic approach to chronic pain management. Future research in this field will likely uncover additional epigenetic mechanisms and pave the way for targeted interventions to alleviate chronic pain and improve the quality of life for affected individuals.

\bibliographystyle{apalike}
\bibliography{references}
\end{document}